\documentclass[12pt]{article}
\usepackage{color}

\bibliography{plain}

\title{\bf Integrable and superintegrable systems with higher order integrals of motion: master function formalism}
\author{ Z. Alizadeh \thanks{ (E-mail:zeinabalizadeh@phd.guilan.ac.ir)} ,
 H. Panahi \thanks{Corresponding author
(E-mail:t-panahi@guilan.ac.ir)}
   \\
{\small Department of Physics ,
University of Guilan , Rasht 51335-1914, Iran.} \\
 }\pagebreak
\begin{document}

\maketitle
\begin{abstract}

 In this article, we construct two-dimensional integrable and superintegrable systems in terms of the master function formalism and relate them to $Mielnik^{,}s$ and $Marquette^{,}s$ construction in supersymmetric quantum mechanics. For two different cases of the master functions, we obtain two different two-dimensional superintegrable systems with higher order integrals of motion.
 \end{abstract}
\pagebreak \vspace{7cm} \pagebreak \vspace{7cm}
\section{Introduction}

  It is known from classical and quantum mechanics that a system
 with N degrees of freedom is called completely integrable if it allows N
 functionally independent constants of the motion[1]. From the
 mathematical and physical point of view, these systems play a
 fundamental role in description of physical systems due to their
 many interesting properties. A system is superintegrable if one could obtain more than N
 constants of the motion and if there exist 2N-1 constants of the motion, the system
 is maximally superintegrable or just superintegrable  provided
 that  the commutator of operators as constants of the motion be
 zero with Hamiltonian of the system[2-8]. Recently the study of
 superintegrable systems has been considered for different
 potentials  and many researches have been
 studied for calculating of the spectrum
 of these systems by different methods.
 In Refs.[9,10], the spectrum of these systems has been calculated by an
 algebraic method  using the realization of some Lie groups.\\
 For a two-dimensional quantum integrable system with  Hamiltonian $H$, there is
 always one operator like $ A_{1}$ which can be commutated with
 Hamiltonian of the system i.e. $
 [H,A_{1}]=0$. For a quantum superintegrable system, one should define another operator
 such as $ A_{2}$ which commutates with the Hamiltonian of system i.e. $
 [H,A_{2}]=0,$ but $[A_{1},A_{2}]\neq0 $. In other words,
 for a two-dimensional superintegrable system, there are two
 integrals of the motion $ (A_{1},A_{2})$ in addition to the
 Hamiltonian.
 The superintegrability with the second and third order integrals was the object
 of a series of articles [11-17]. The systems studied have second and third order integrals. Although superintegrability and supersymmetric quantum mechanics (SUSYQM) are two separated fields,  many quantum systems such as the harmonic oscillator, the Hydrogen atom and the Smorodinsky-Winternits potential , have both supersymmetry and superintegrable conditions[18-23]. These articles show that superintegrability is accurately connected with supersymmetry. For example, in Ref. [24], Marquette  used  the results obtained by Mielnik [25] and generated new superintegrable systems.  Mielnik has shown that the factorization  of
 second order operators is not essentially unique. He has considered the Hamiltonian of the
 harmonic oscillator in one dimension as the simplest case: \begin{equation}
 H=-\frac{1}{2}\frac{d^{2}}{dx^{2}}+\frac{1}{2}{x^{2}},\end{equation}
 where can be factorized by two-type of the first order operators of creation and annihilation  as follows: \begin{equation}
 a_{\pm}=\frac{1}{\sqrt{2}}(\mp\frac{d}{dx}+x),\;\;\;\;\;
 b_{\pm}=\frac{1}{\sqrt{2}}(\mp\frac{d}{dx}+\beta(x)).\end{equation}
{For two superpartner Hamiltonians $H_1$ and $H_2$ where $a_+a_-=H-\frac{1}{2}=H_1$ and $a_-a_+=H+\frac{1}{2}=H_2$, he has demanded that  $ H_{2}=b_{-}b_{+} $ and  obtained
 the inverted product $b_{+}b_{-} $ as a certain new Hamiltonian: \begin{equation}
 H^{\prime}=b_+b_-=-\frac{1}{2}\frac{d^{2}}{dx^{2}}+\frac{x^{2}}{2}-\varphi^{\prime}(x),\end{equation}
 where $ \varphi(x)$ is a function obtained from the general solution of Riccati equation considering $\beta=x+\varphi(x)$ .
 The creation and annihilation operators of the third order for $H^{\prime}$ are described by  expressions $  s_{+}=b_{+}a_{+}b_{-}$, $ s_{-}=b_{+}a_{-}b_{-}$, that $a_{+}$ and $a_{-}$ are the
 creation and annihilation operators for $H _{2}$. Marquette [24]  has taken  the Hamiltonian $H_{2}$ in the $x$-axis and  its superpartner $H^{\prime}$  given by Eq.(3) in the $y$-axis. Hence he has obtained a two-dimensional superintegrable
 system as $H_{s}=H_{x}+H_{y}$, where can be separated in  Cartesian coordinates with creation and annihilation operators $ {}a_{+}(x)$, $a_{-}(x)$, $s_{+}(y)$ and $s_{-}(y)$.
 Also, he has shown that the Hamiltonian $H_{s}$  possesses the following integrals of motion
 \begin{eqnarray}
 \mathcal{K}=\mathrm{H_{x}-H_{y}},\nonumber\\ \mathcal{A}\mathrm{}_ {1}= \textit{a}_{+}(x)s_{-}(y)-\textit{a}_{-}(x)s_{+}(y),\nonumber\\
 \; \mathcal{}A\mathrm{}_ {2}=\textit{a}_{-}(x)s_{+}(y)+\textit{a}_{+}(x)s_{-}(y),\end{eqnarray}
 where these integrals are of order
 2, 3 and 4 for shape invariant potentials[24].\\
 On the other hand, in Refs. [26,27], the authors have shown that the second-order differential
 equations and their associated differential equations in mathematical physics have the
 shape invariant property of supersymmetry quantum mechanics. They have  shown that by using  a polynomial
 of a degree not exceeding two, called the master function, the associated
 differential equations can be factorized into the product of rising and lowering operators. The master function formalism has been used in relativistic quantum mechanics for solving the Dirac equation [28,29].
 \\  As  $Mielnik^{,}s$- $Marquette^{,}s$ method for generating superintegrable systems
can be applied to other systems obtained in the context of supersymmetric quantum mechanics hence  in this paper, we  show that the supersymmetry method for obtaining the integrable and superintegrable systems can be related to master function formalism.
 In fact, we use the master function approach for 1-demensional shape invariant potentials and generate 2-dimensional integrable systems.  Also for a particular class of shape invariant systems, we generate 2-dimensional supperintegrable systems. This class contains the harmonic oscillator, the singular harmonic oscillator and their suppersymmetric isospectral deformations. \\
 The paper is presented as follows: in section 2,  we review how one can generate integrals of motion for two-dimensional superintegrable system from the creation and annihilation operators.
 In section 3, we consider a particular quantum system for applying the $Mielnik^{,}s$- $Marquette^{,}s$ method and obtain a superintegrable potential separable in cartesian coordinates.
 In section 4, we  briefly review the master function formalism and then in section 5, we use this
 approach to obtain integrable systems and particular cases of the superintegrable systems that satisfy the oscillator-like (Heisenberg) algebra with higher order integrals
 of motion in terms of the master function and weight function. In section 6, we give two examples to show how this method works in constructing oscillator-like two-dimensional superintegrable systems. Paper ends with a brief conclusion in section 7.

 \section{  Two-dimensional superintegrable system and its integrals of motion  }
According to Refs. [24,30,31], for a two-dimensional Hamiltonian separable in Cartesian coordinates as:
 \begin{equation}
 H(x,y,p_{x},p_{y})=H_{x}(x,p_{x})+H_{y}(y,p_{y}),
 \end{equation}
 where the creation and annihilation operators (polynomial in momenta) $A_{+}(x)$, $A_{-}(x)$, $A_{+}(y)$ and $A_{-}(y)$  satisfy the following equations
 \begin{eqnarray}
 [H_{x},A_{-}(x)]=-\lambda_{x} A_{-}(x),\;\;\;\;\;\;[H_{y},A_{-}(y)]=-\lambda_{y} A_{-}(y),\nonumber\\\;
 [H_{x},A_{+}(x)]=\lambda_{x} A_{+}(x),\hspace{0.8cm}\;\;\;\;\; [H_{y},A_{+}(y)]=\lambda_{y} A_{+}(y),
 \end{eqnarray}
 one can show that the  operators  $f_{1}=A_+^{m}(x)A_-^{n}(y)$ and $f_{2}=A_-^{m}(x)A_+^{n}(y)$  commute with the Hamiltonian H, that is
\begin{equation}
 [H,f_{1}]=[H,f_{2}]=0,
 \end{equation}
 if\begin{equation}
 m\lambda_{x}-n\lambda_{y}=0,\;\;\;\;\;\;\;\;\ m,n\in Z^{+}.
 \end{equation}
Also the following sums of $f_1$ and $f_2$  commute with the Hamiltonian
 \begin{equation}
 I_{1}=A_{+}^{m}(x)A_{-}^{n}(y)-A_{-}^{m}(x)A_{+}^{n}(y),\;\;\;\;\;\;\
 I_{2}=A_{+}^{m}(x)A_{-}^{n}(y)+A_{-}^{m}(x)A_{+}^{n}(y),
 \end{equation}
that is, $I_1$ and $I_2$ are  the integrals of motion.  The order of these integrals of motion  depends on the order of the creation and annihilation operators.  On the other hand, the Hamiltonian $H$ possesses a second order integral as  $K=H_{x}-H_{y}$, such that the integral $I_2$ is the commutator of $I_1$ and $K$. Thus the  Hamiltonian $H$ is a  superintegrable system and $H$, $I_{1}$ and $K$  are its integrals of motion.

  \section{  Mielnik- Marquette method and superintegrable model obtained from shifted oscillator Hamiltonian }
 In this section, for reviewing of the Mielnik-Marquette method, we consider shifted oscillator Hamiltonian as
 \begin{equation}
 H=-\frac{d^{2}}{dx^{2}}+\frac{1}{4}\omega^{2}(x-\frac{2b}{\omega})^{2}-\frac{\omega}{2}. \end{equation}
 We introduce the following first order operators
 \begin{equation}
 a_{-}=\frac{d}{dx}+\frac{1}{2}\omega x-b,\;\;\;\;\;\;\ a_{+}=-\frac{d}{dx}+\frac{1}{2}\omega x-b,
 \end{equation}
 where the supersymmetric partner Hamiltonians are calculated as
 \begin{eqnarray}
 H_{1}=a_{-}a_{+}=H+\omega,\;\;\;\;\;
 H_{2}=a_{+}a_{-}=H.
 \end{eqnarray}
 It is obvious that $H_{1}$ and $H_{2}$ have the shape invariant properties. Now, according to Eq.(2), we define the new operators $b_{-}$ and $b_{+}$ such that
 \begin{equation}
 H_{1}=H+\omega=b_{-}b_{+}.
 \end{equation}
 The above equation gives the following Riccati equation
 \begin{equation}
 \beta^{2}+\beta^{\prime}=\frac{1}{4}\omega^{2}x^{2}-b\omega x+b^{2}+\frac{\omega}{2},
 \end{equation}
 where a particular solution is
 \begin{equation}
 \beta(x)=\beta_{0}(x)=\frac{1}{2}\omega x-b.
 \end{equation}
 Now, if we consider
 \begin{equation}
 \beta(x)=\beta_{0}(x)+\varphi(x),
 \end{equation}
 then we can obtain the following first order linear inhomogeneous equation
 \begin{equation}
 z^{\prime}+(-2\beta_{0})z=1,
 \end{equation}where $z=\frac{1}{\varphi(x)}$. After solving the above equation, we get
 \begin{equation}
 \varphi(x)=\frac{1}{z(x)}=\frac{e^{-\frac{\omega}{2}x^{2}+2bx}}{\sqrt{\frac{\pi}{2\omega}}e^{\frac{2b^{2}}
 {\omega}}Erf(\sqrt{\frac{\omega}{2}}(x-\frac{2b}{\omega})+C},
 \end{equation}where C is the constant of integration.
 Using the function $\varphi(x)$, we obtain
 \begin{equation}
 H^{\prime}= b_+b_- =H_{1}-\varphi^{\prime}(x),
 \end{equation}
 where its creation and annihilation operators  are given by following expressions
 \begin{equation}
 s_{+}=b_{+}a_{+}b_{-},\;\;\;\;\;\;\;\ s_{-}=b_{+}a_{-}b_{-}.
 \end{equation}
 According to Marquette method, we take the $x$ axis for Hamiltonian $H_{1}$ and the $y$ axis for its superpartner $H^{\prime}$ and we have the following two-dimensional superintegrable system
 \begin{eqnarray}
 H_{s}=H_{x}+H_{y}\hspace{10.5cm}\nonumber\\=H_{1}+H^{\prime}=-\frac{d^{2}}{dx^{2}}-\frac{d^{2}}{dy^{2}}
 +\frac{1}{4}\omega^{2}(x-\frac{2b}{\omega})^{2}+\frac{1}{4}\omega^{2}(y-\frac{2b}{\omega})^{2}
 -\frac{\omega}{2}-\frac{d\varphi}{dy}.
 \end{eqnarray}
 This Hamiltonian possesses the integral of motion  given by Eq.(4), which are of order 2, 3 and 4.

  \section{The Master function formalism }
 According to Refs. [26,27], the general form of the differential equation in master function approach is written as:
 \begin{equation}
 A(x)\Phi_{n}^{\prime\prime}+\frac{(A(x)w(x))^{\prime}}{w(x)}\Phi_{n}^{\prime}(x)
 -\left(n(\frac{(A(x)w(x))^{\prime}}{w(x)})^{\prime}+\frac{n(n-1)}{2}A^{\prime\prime}(x)\right)\Phi_{n}(x)=0,
 \end{equation}
 where $A(x)$ as master function is at most a second order polynomial and $w(x)$ is the non-negative weight function in interval $(a,b)$.
 By differentiating Eq. (22) $m$ times and then multiplying it by
 $ (-1)^{m}A^{\frac{m}{2}}(x)$, we get the following associated second-order differential
 equation in terms of the master function and weight function
 \begin{eqnarray}
 A(x)\Phi_{n}^{\prime\prime}+\frac{(A(x)w(x))^{\prime}}{w(x)}\Phi_{n}^{\prime}(x)+
 [-\frac{1}{2}(n^{2}+n-m^{2})A^{\prime\prime}+
 (m-n)(\frac{A(x)w^{\prime}(x)}{w(x)})^{\prime}
 \nonumber\\-\frac{m^{2}}{4}\frac{(A^{\prime}(x))^{2}}{A(x)}-
 \frac{m}{2}\frac{A^{\prime}(x)w^{\prime}(x)}{w(x)}]\Phi_{n,m}(x)=0,\end{eqnarray}
 where
 \begin{equation}
 \Phi_{n,m}(x)=(-1)^{m}A^{\frac{m}{2}}(\frac{d}{dx}^{m})\Phi_{n}(x).\end{equation}
 Changing the variable $\frac{dx}{dr}=\sqrt{A(x)}$, and defining the
 new function $\Psi_{n}^{m}(r)=A^{\frac{1}{4}}(x)w^{\frac{1}{2}}(x)\phi_{n,m}(x)$,
 one can obtain the Schrodinger equation as:
 \begin{equation}
 -\frac{d^{2}}{dr^{2}}\Psi_{n}^{m}(r)+v_{m}(x(r))\Psi_{n}^{m}(r)=E(n,m)\Psi_{n}^{m}(r),\;\;\;\;\;\;\;
 \;m=0,1,2,..., \end{equation}
 where the most general shape invariant potential is:
  \begin{eqnarray}
 v_{m}(x(r))=-\frac{1}{2}\left(\frac{A(x)w^{\prime}(x)}{w(x)}\right)^{\prime}
 -\frac{2m-1}{4}A^{\prime\prime}(x)\nonumber\\+\frac{1}{4A(x)}\left(\frac{A(x)w^{\prime}(x)}{w(x)}\right)^{2}+
 \frac{m}{2}\frac{A^{\prime}(x)w^{\prime}(x)}{w(x)}+\frac{4m^{2}-1}{16}\frac{A^{\prime2}(x)}{A(x)},
 \end{eqnarray}
 and the energy spectrum $ E(n,m)$ is as:
 \begin{equation}
 E(n,m)=-(n-m+1)\left[\left(\frac{A(x)w^{\prime}(x)}{w(x)}\right)^{\prime}+\frac{1}{2}(n+m)A^{\prime\prime}(x)\right].\end{equation}
 According to Refs. [26,27] the first-order deferential operators are written as:
 \begin{equation}
 A_{\pm}=\mp\frac{d}{dr}+W_{m}(x(r)), \end{equation}
 where the superpotential $W_{m}(x(r))$ is expressed in terms of the master function $A(x)$ and weigh function $w(x)$ as:
 \begin{equation}
 W_{m}(x(r))=-\frac{A(x)w^{\prime}(x)/2w(x)+((2m-1)/4)A^{\prime}(x)}{\sqrt{A(x)}}.\end{equation}
 The Hamiltonian $H_{1}$ and $H_{2}$  called the superpartner Hamiltonians are written as
 \begin{eqnarray} H_{1}=A_{-}A_{+}=-\frac{d^{2}}{dr^{2}}+W_{m}^{2}(r)+W_{m}^{\prime}(r)=
 -\frac{d^{2}}{dr^{2}}+v_{1}(r),\nonumber\\
 H_{2}=A_{+}A_{-}=-\frac{d^{2}}{dr^{2}}+W_{m}^{2}(r)-W_{m}^{\prime}(r)=-\frac{d^{2}}{dr^{2}}+v_{2}(r),
 \end{eqnarray}
 where $v_{1}(r)$ and $v_{2}(r)$ are called the partner potentials in the concept of supersymmetry in nonrelativistic quantum mechanics. Furthermore, if the partner potentials have the same shape and differ only in parameters, then potentials  $v_{1}(r)$ and $v_{2}(r)$ are called the shape invariant potentials that satisfy in
 \begin{equation}
 v_{1}(r,a_{0})=v_{2}(r,a_{1})+R(a_{1}), \end{equation}
 where $R(a_{1})$ is independent of any dynamical variable and $a_{1}$ is a function of $a_{0}$. Potentials which satisfy in this condition are exactly solvable, although shape invariance is not the most general integrability or   superintegrability condition.

  \section{Integrable and superintegrable systems obtained from the master function formalism }
 In this section, we try to relate the Mielnik-Marquette method to the master function approach. Hence we define the following new operators:
 \begin{equation}
 B_{\pm}=\mp\frac{d}{dr}+\omega(r), \end{equation}
 where $\omega(r)$ as the new superpotential must be related to the general form of the master function superpotential $W_{m}(x(r))$.
  Their product yields to  Hamiltonians as:
 \begin{eqnarray}
 B_{-}B_{+}=-\frac{d^{2}}{dr^{2}}+\omega^{2}(r)+\omega^{\prime}(r),\nonumber\\
 B_{+}B_{-}=-\frac{d^{2}}{dr^{2}}+\omega^{2}(r)-\omega^{\prime}(r).\end{eqnarray}
 Now if we demand $A _{-}A_{+}= B _{-}B_{+}$ then we can obtain the following Riccati equation in terms
 of master function:
 \begin{equation}
 \omega^{2}(r)+\omega^{\prime}(r)=W_{m}^{2}(r)+W_{m}^{\prime}(r), \end{equation}
 where a particular solution is $\omega(r)=W_{m}(r)$.
 The general solution can be obtained like :
 \begin{equation}
 \omega(r)=W_{m}(r)+\lambda(r), \end{equation}
 which yields: \begin{equation}
 \lambda^{2}(r)+2W_{m}(r)\lambda(r)+\lambda^{\prime}(r)=0. \end{equation}
 We consider the transformation $f(r)=\frac{1}{\lambda(r)}$
 and obtain a first order linear inhomogeneous differential equation as:
 \begin{equation}
 f^{\prime}(r)-2W_{m}(r)f(r)=1, \end{equation}
 which the general solution is :
 \begin{equation}
 f(r)=\exp \left[2\int W_{m}(r)dr \right]\left(C+\int\exp\left[{2}\int W_{m}(r^{\prime})dr^{\prime}\right]dr \right ), \end{equation}
 where C is constant. Hence:
{\begin{equation}
 \omega(r)=W_{m}(r)+\frac{e^{-\int2W_{m}(r)dr}}{C+\int e^{\int2W_{m}(r^{\prime})dr^{^{\prime}}}dr}. \end{equation}
  Using the function $f(r)$ given by (38), the superpartner Hamiltonian is given by:
  \begin{equation}
 H^{\prime}=H_{2}-\lambda^{\prime}(r)=-\frac{d^{2}}{dr^{2}}+W_{m}^{2}(r)-
 W_{m}^{\prime}(r)-\lambda^{\prime}(r), \end{equation}
 which is the general form of Hamiltonian in terms
 of master function. Now if we catch $H_{r}=H_{2}$ and $H_{r^{\prime}}=H^{\prime}$ (the Hamiltonian
 $H^{\prime}$ is thus given in terms of the variable $ r ^{\prime}$ vertical to r) then we obtain  a new two-dimensional integrable Hamiltonian as:
 \begin{equation}
 H_{s}=H_{r}+H_{r^{\prime}}=-\frac{d^{2}}{dr^{2}}-\frac{d^{2}}{dr^{\prime^{2}}}+W_{m}^{2}(r)-W_{m}^{2}
 (r^{\prime}) +W_{m}^{\prime}(r)-W_{m}^{\prime}(r^{\prime})-\lambda^{\prime}(r^{\prime}).\end{equation}
 Therefore we have obtained the general form of the 2-dimensional integrable Hamiltonian in terms of master function in which can be separated in radial coordinates. This separation of variable implies the existence of a second order integral as $\mathcal{}K\mathrm{}=H_{r}-H_{r^{\prime}}$. Hence, $H_{s}$ is a integrable system. Now, for generating superintegrable systems,  we can obtain  the creation and annihilation operators for $H^\prime$ from $H_{2}$ as
 \begin{equation}
 S_{+}=B_{+}A_{+}B_{-},\;\;\;\;\;\;S_{-}=B_{+}A_{-}B_{-}, \end{equation}
where $A_{\pm}$ and $B_{\pm}$ were given in  Eqs. (28),(32). As these ladder operators satisfy the relation  given by (6) only for a particular class of shape invariant systems so in general form, the 2-dimensional system  $H_{s}$,  obtained from a given master function, is not a superintegrable system. In fact, this class contains the harmonic oscillator,
the singular harmonic oscillator and their suppersymmetric iso-spectral deformations.  \\
Hence if it exists,  according to Eq. (9) we can  obtain the  integrals of motion for Hamiltonian (41) as
 \begin{eqnarray}
 \mathcal{}K=\mathrm{}H_{r}-H_{r^{\prime}},\nonumber\\
 \mathcal{}A\mathrm{}_ {1}=A_{+}^{m}(r)S_{-}^{n}(r^{\prime})-A_{-}^{m}(r)S_{+}^{n}(r^{\prime}),\nonumber\\
 \mathcal{}A\mathrm{}_ {2}=A_{+}^{m}(r)S_{-}^{n}(r^{\prime})+A_{-}^{m}(r)S_{+}^{n}(r^{\prime}).
 \end{eqnarray}
  In the next section, we apply this formalism for some particular cases of shape invariant potentials in terms of master function.

  \section{Examples of two-dimensional superintegrable systems as a result of master function approach}
 In this section,  we would apply the master function formalism of the previous section for two examples and show how these results allow us to obtain 2-dimensional superintegrable systems with higher order integrals.\\
 \textbf{Example 1}\\
 Let $ A(x)=1 $, then according to Ref. [26] $, w(x)=e^{-\frac{\beta}{2}x^{2}}$ that   $ x=r-\frac{2\alpha}{\beta}$, $\beta>0$ and the interval is $(-\infty,+\infty)$. Using Eq.(29), we obtain the superpotential as:
 \begin{equation}
 W_{m}(r)=\frac{\beta}{2}(r-\frac{2\alpha}{\beta}). \end{equation}
 According to Eq. (27), the energy spectrum is as
 \begin{equation}
 E=n-m+1, \end{equation}
 also the ladder operators given by equation (28) related to Eq. (44), satisfy a Heisenberg algebras (6). Now, substituting expression $W_{m}(r)$ in Eq.(38) yields the following relation in terms of the Error function: \begin{equation}
 \lambda(r)=\frac{e^{\beta\frac{r^{2}}{2}+\alpha r}}{C+\sqrt{\frac{\pi}{2\beta}}e^{\frac{2\alpha^{2}}{\beta}}Erf\left(\sqrt{\frac{\beta}{2}}
 (r-\frac{2\alpha}{\beta})\right)},\end{equation}
 and so
 \begin{equation}\omega(r)=W_{m}(r)+\lambda(r)=\frac{\beta}{2}(r-\frac{2\alpha}{\beta})
 +\frac{e^{\beta\frac{r^{2}}{2}+\alpha r}}{C+\sqrt{\frac{\pi}{2\beta}}e^{\frac{2\alpha^{2}}{\beta}}Erf\left(\sqrt{\frac{\beta}{2}}
 (r-\frac{2\alpha}{\beta})\right)}. \end{equation}
  Substituting this expression in Eqs. (40),(41), yield the family of superpartner $ H^{\prime}$ and a two-dimensional superintegrable Hamiltonian respectively as:
 \begin{eqnarray}
 H^{\prime}=H_{2}-\lambda^{\prime}(r),\nonumber\\H_{s}=-\frac{d^{2}}{dr^{2}}-\frac{d^{2}}{dr^{\prime2}}+
 \frac{\beta^{2}}{4}(r-r^{\prime})(r+r^{\prime}-\frac{4\alpha}{\beta})-\lambda^{\prime}(r^{\prime}), \end{eqnarray}
 where
 \begin{equation}\lambda^{\prime}(r^{\prime})=\frac{(\beta r^{\prime}+\alpha) e^{\frac{\beta}{2}r^{\prime 2}+\alpha r^{\prime}}}{C+\sqrt{\frac{\pi}{2\beta}}e^{\frac{2\alpha^{2}}{\beta}}Erf\left(\sqrt{\frac{\beta}{2}}
 (r^{\prime}-\frac{2\alpha}{\beta})\right)}-\frac{(e^{\frac{\beta}{2}r^{\prime 2}+\alpha r^{\prime}})(e^{\frac{2\alpha^{2}}{\beta}}e^{-\frac{1}{2}\beta(r^{\prime}-\frac{2\alpha}{\beta})^{2}})}
 {\left[C+\sqrt{\frac{\pi}{2\beta}}e^{\frac{2\alpha^{2}}{\beta}}
 Erf\left(\sqrt{\frac{\beta}{2}} (r^{\prime}-\frac{2\alpha}{\beta})\right)\right]^{2}}.\end{equation}
 and
 \begin{equation}
 H_{2}=-\frac{d^{2}}{dr^{2}}+\frac{\beta^{2}}{4}(r-\frac{2\alpha}{\beta})^{2}-\frac{\beta}{2}.\end{equation}
 We can find the general form of the operators $S_{+}$ and $S_{-}$ in terms of the master function for this oscillator-like potentials as follows:
 \begin{eqnarray}
 S_{+}=-\frac{d^{3}}{dr_{3}}-W_{m}\frac{d^{2}}{dr^{2}}+(-2\omega^{\prime}-W_{m}^{\prime}+\omega^{2}
 )\frac{d}{dr}+(-\omega^{\prime\prime}-W_{m}^{\prime}\omega-W_{m}\omega^{\prime}+\omega\omega^{\prime}
 + W_{m}\omega^{2}),\nonumber\\S_{-}=\frac{d^{3}}{dr_{3}}-W_{m}\frac{d^{2}}{dr^{2}}+(-2\omega^{\prime}-
 W_{m}^{\prime}-\omega^{2} )\frac{d}{dr}+(\omega^{\prime\prime}-W_{m}^{\prime}\omega-W_{m}\omega^{\prime}+
 \omega\omega^{\prime}+ W_{m}\omega^{2}).\hspace{.8cm}\end{eqnarray}
 Thus we have obtained a 2-dimensional superintegrable system with integrals given by Eq.(43) as:
 \begin{eqnarray}
 \mathcal{}K=\mathrm{}H_{r}-H_{r^{\prime}},\nonumber\\
 \mathcal{}A\mathrm{}_ {1}=A_{+}(r)S_{-}(r^{\prime})-A_{-}(r)S_{+}(r^{\prime}),\nonumber\\
 \mathcal{}A\mathrm{}_ {2}=A_{+}(r)S_{-}(r^{\prime})+A_{-}(r)S_{+}(r^{\prime}),
 \end{eqnarray}
 where
 \begin{eqnarray}\mathcal{}K=\mathrm{}- \frac{d^{2}}{dr^{2}}+\frac{d^{2}}{dr^{\prime2}}+W_{m}^{2}(r)-W_{m}^{2}(r^{\prime})+
 W_{m}^{\prime}(r)+W_{m}^{\prime}(r^{\prime})+\lambda^{\prime}(r^{\prime}),\nonumber\\
 \mathcal{}A\mathrm{}_{1}=2W_{m}(r^{\prime})
 \frac{d^{3}}{drdr^{\prime2}}-2W_{m}(r)\frac{d^{3}}{dr^{\prime3}}+
 2W_{m}^{\prime}(r^{\prime})\frac{d^{2}}{drdr^{\prime}}
 +(2W_{m}^{\prime}(r^{\prime})\omega(r^{\prime})\nonumber\\
 +2W_{m}(r^{\prime})\omega^{\prime}(r^{\prime})-
 2\omega^{2}(r^{\prime}) W_{m}(r^{\prime})\frac{d}{dr}+W_{m}(r)\nonumber\\(-4\omega^{\prime}(r^{\prime})+
 2\omega^{2}(r^{\prime}))\frac{d}{dr^{\prime}}
 W_{m}(r)(2\omega\omega^{\prime}-2\omega^{\prime\prime}),\\
 \mathcal{}A\mathrm{}_{2}=2\frac{d^{4}}{drdr^{\prime3}}+(W_{m}(r^{\prime})-W_{m}(r))\frac{d^{3}}
 {drdr^{\prime2}}+(4\omega^{\prime}(r^{\prime})-
 2\omega^{2}(r))\frac{d^{2}}{drdr^{\prime}}\nonumber\\-
 2W_{m}(r)W_{m}(r^{\prime})
 \frac{d^{2}}{dr^{\prime2}}-2W_{m}(r)W_{m}^{\prime}(r^{\prime})
 \frac{d}{dr^{\prime}} +(2\omega^{\prime\prime}
 (r^{\prime})\nonumber\\-2\omega(r^{\prime})\omega^{\prime}(r^{\prime}))\frac{d}{dr}+
 W_{m}(r)(2\omega^{2}(r^{\prime})
 W_{m}(r^{\prime})\nonumber\\-2W_{m}(r^{\prime})\omega^{\prime}(r^{\prime})-2W_{m}^{\prime}
 (r^{\prime})\omega(r^{\prime})).\nonumber \end{eqnarray}
 These integrals are of order 2, 3 and 4. \\\\
 \textbf{ Example 2}\\
 According to Ref. [26] for $A(x)=x$, we have $w(x)=x^{\alpha}e^{-\beta x}$, $x=\frac{r^{2}}{4},$ $\alpha>-1$, $\beta>0$  and the interval is $ [0,+\infty)$. Now, using Eq.(29), the superpotential and  the energy spectrum are as
 \begin{equation}
 W_{m}(r)=-\frac{1}{r}(\alpha+m-\frac{1}{2})+\frac{\beta}{4}r,\;\;\;\;\; E=\beta (n-m+1).\end{equation}
  This system  has also the ladder operators that satisfy the form of  Eq. (6) hence
substituting expression $W_{m}(r)$ in Eq.(38) yields the following relation in terms of Whittaker function:
 \begin{eqnarray}
 f(r)=\frac{\beta^{-(\alpha+m)}}{(\alpha+m)}\left(\frac{r}{2}\right)^{-(2\alpha+2m-1)}e^{\frac{1}{4}\beta r^{2}}\left (C+
 {e^{-\frac{1}{8}\beta r^{2}}}\left[\frac{1}{(\alpha+m+1)}\left(\frac{\beta r^{2}}{4}\right)^{\frac{\alpha}{2}+\frac{m}{2}} \right.\right. \nonumber\\ \left. \left.
 M_{\frac{1}{2}\alpha+\frac{1}{2}m,\frac{1}{2}\alpha+\frac{1}{2}m+\frac{1}{2}}
 ( \frac{1}{4}\beta r^{2}) +\left(\frac{\beta r^{2}}{4}\right)^{\frac{\alpha}{2}+\frac{m}{2}-1}{M_{\frac{1}{2}\alpha+
 \frac{1}{2}m+1,\frac{1}{2}\alpha+\frac{1}{2}m+\frac{1}{2}}(\frac{1}{4}\beta r^{2})}\right] \right),\end{eqnarray}
 where the Whittaker function ${M_{\mu,\nu}(z)}$ is the solution of the following differential equation: \begin{equation}y^{\prime\prime}+(-\frac{1}{4}+\frac{\mu}{z}+\frac{\frac{1}{4}-\nu^{2}}{z^{2}}
 )y=0 .\end{equation}
 It can be  also defined in terms of  the confluent hypergeometric function as:
 \begin{equation}{M_{\mu,\nu}(z)}=e^{(-\frac{1}{2}z)}z^{(\frac{1}{2}+\nu)} \;_{1}
 F_{1}(\frac{1}{2}+\nu-\mu,1+2\nu,z).\end{equation}
  The family of superpartner Hamiltonians $H^{\prime}$ and the two-dimensional superintegrable Hamiltonian $H_{s}$ are thus calculated by Eqs. (40), (41) respectively.
 The creation and annihilation operators for the Hamiltonian $H_{2}$ are as
 \begin{equation}M_{+}(r)=A_{+}^{2}(r)A_{-}(r),\;\;\;\;\;\;\ M_{-}(r)=A_{+}(r)A_{-}^{2}(r),\end{equation}
 where $A_{\pm}(r)$ is given in Eq. (28)and from Eq.(42), we have  the creation and annihilation operators of the Hamiltonian $H^{\prime}$ as:
\begin{equation}R_{+}(r^{\prime})=B_{+}(r^{\prime})M_{+}(r^{\prime})B_{-}(r^{\prime})
 ,\;\;\;\;\;\;\ R_{-}(r^{\prime})=B_{+}(r^{\prime})M_{-}(r^{\prime})B_{-}(r^{\prime}),\end{equation}
 where $B_{\pm}(r)$ is given by (32)and (39).
 We can also find  the integrals of motion of the Hamiltonian $H_{s}$ from Eq.(43) as:
 \begin{eqnarray}
 \mathcal{}K=\mathrm{}H_{r}-H_{r^{\prime}},\nonumber\\
 \mathcal{}A\mathrm{}_ {1}=M_{+}(r)R_{-}(r^{\prime})-M_{-}(r)R_{+}(r^{\prime}),\nonumber\\
 \mathcal{}A\mathrm{}_ {2}=M_{+}(r)R_{-}(r^{\prime})+M_{-}(r)R_{+}(r^{\prime}),
 \end{eqnarray}
 that are of the order 2, 7 and 8.

   \section{Conclusion}
  In this article, we have shown how the supersymmetric quantum mechanics gives a procedure for constructing  two-dimensional integrable and superintegrable systems with higher order integrals of motion. We have used the results obtained by Mielnik in the concept of SUSYQM and  related  it to master function formalism for constructing   two-dimensional integrable and superintegrable systems. From this procedure, we have generated the superintegrable systems for two different cases of master functions  $A(x)=1$ and $A(x)=x$, and have shown that the higher integrals of motion are in order 2, 3, 4 and 2, 7, 8 respectively.

\end{document}